# Transport spectroscopy of low disorder silicon tunnel barriers with and without Sb implants

A Shirkhorshidian[1,2] , N C Bishop[2], J Dominguez[2], R K Grubbs[2], J R Wendt[2], M P Lilly[2,3] and M S Carroll[2]

[1] University of New Mexico, Albuquerque, New Mexico 87131, USA
[2] Sandia National Laboratories, Albuquerque, New Mexico 87185, USA
[3] Center for Integrated Nanotechnologies, Sandia National Laboratories, Albuquerque, New Mexico 87185, USA

Email: ashirkh@sandia.gov and mscarro@sandia.gov

**Abstract**
We present transport measurements of silicon MOS split gate structures with and without Sb implants. We observe classical point contact (PC) behavior that is free of any pronounced unintentional resonances at liquid He temperatures. The implanted device has resonances superposed on the point contact transport indicative of transport through the Sb donors. We fit the differential conductance to a rectangular tunnel barrier model with a linear barrier height dependence on source-drain voltage and non-linear dependence on gate bias. Effects such as Fowler-Nordheim (FN) tunneling and image charge barrier lowering (ICBL) are considered. Barrier heights and widths are estimated for the entire range of relevant biases. The barrier heights at the locations of some of the resonances for the implanted tunnel barrier are between 15-20 meV, which are consistent with transport through shallow partially hybridized Sb donors. The dependence of width and barrier height on gate voltage is found to be linear over a wide range of gate bias in the split gate geometry but deviates considerably when the barrier becomes large and is not described completely by standard 1D models such as FN or ICBL effects.

## 1. Introduction

Tunnel barriers formed using electrostatic gate structures are common experimental platforms for probing one-dimensional systems [1,2] including performing single impurity transport spectroscopy [3,4]. They are also an essential building block for zero-dimensional lateral quantum dots [5]. Transport near equilibrium in quantum point contacts has been examined in great depth in model material systems such as GaAs. Silicon split-gate tunnel barriers are increasingly utilized because of recent successes with forming silicon qubits using laterally gated electrostatic quantum dots [6] as well as for donor spectroscopy. Despite recent successes, a tunnel barrier in MOS without unintentional resonances, i.e. a clean barrier, has been challenging to form and there is little reported directly about the characterization of split gate barrier height, width and their dependence on voltage for the MOS system.

A common transport spectroscopy measurement is to monitor the current through a constriction while varying the gate and source-drain bias. The gate bias changes the barrier height and width. The source-drain bias changes the relative alignment of the Fermi level to the barrier height and bound state energy levels within the barrier thereby revealing resonances in the energy space. Improved understanding of how to identify barrier height and width and map this to donor spectroscopy is desirable. MacLean et al. developed a model that assumes a linear dependence of barrier height on gate and source-drain bias [7] and that fits experimental results for split gate tunnel barriers in GaAs and SiGe/sSi [8], at least for limited bias regimes. This linear dependence model could be used for this purpose. However, it is unclear how well this model fits a tunnel barrier's conductance dependence on different gate voltages over a wider range of bias typically used for donor spectroscopy and whether it can be used to extract useful information for transport spectroscopy when additional resonances (i.e., donor atoms) have been injected into the tunnel barrier. In addition, understanding the role of high fields and geometry are important for the voltage ranges and split-gate geometry used for this form of single-shallow-bound-impurity spectroscopy [9].

In this paper, we fabricate, measure and compare MOS split gate tunnel barriers with and without Sb implants formed with processing parameters similar to those used in CMOS processing. Clean barrier behavior is observed in the un-implanted case for which process details are provided in this letter. The implanted barrier has numerous resonances owing to single shallow impurities near the surface that short-circuit the tunnel barrier. A rectangular tunnel barrier model that assumes a linear barrier height dependence on source-drain and gate voltage is used to estimate barrier heights and widths for the entire range of relevant biases in the split-gate geometry. Fowler-Nordheim and image charge barrier lowering are also considered. The barrier heights at the locations of some of the resonances for the implanted tunnel barrier are between 15-20 meV, which are consistent with transport through shallow partially hybridized Sb donors as predicted by theory and observed in previous work for other impurities in Si (i.e., P and As). Sb is of interest for quantum computing because it offers improved positional control for single impurity devices formed by ion implantation, however, it has been less well studied than P or As from the perspective of activation, residual damage and transport spectroscopy in a split gate geometry. Furthermore, we observe that the voltage dependence of barrier height and width is found to be linear over a wide range of gate bias but deviates considerably when the barrier becomes large.

## 2. Methods

The devices studied in this paper are electrostatically formed tunnel barriers in silicon using a surface Al gate and poly-Si depletion gates. The depletion gates are inserted below the Al gate to form a PC configuration. The devices begin with a p-type Si substrate with $10^{15}$ cm$^{-3}$ boron. Arsenic is implanted to form the n$^+$ source and drain contacts. A 35 nm silicon dioxide (SiO$_2$) gate dielectric is subsequently formed using thermal oxidation. Degenerately n-doped poly-Si is deposited over the SiO$_2$ layer and patterned using electron beam lithography (EBL) and a dry etch. A scanning electron microscope (SEM) image of the poly-Si gate pattern of a representative device is shown in figure 1(a).

We characterize a PC (sample A) that is not implanted and an Sb implanted PC (sample B). For sample B, a second EBL step follows the poly-Si etch. An 80 nm × 80 nm implantation window is created between the top plunger (TP) and center plunger (CP) poly-Si gates into which Sb is implanted with an energy and dose of 120 keV and $4 \times 10^{11}$ cm$^{-2}$, respectively. This corresponds to an average of ten Sb donors in the constriction at a depth of approximately 27 nm below the Si/SiO$_2$ interface.

Both samples then go through a second thermal oxidation at 900°C for 24 minutes so that an additional 10-20 nm of SiO$_2$ is grown on the poly-Si gates. The reoxidation step serves many purposes that include activating the Sb, repairing the damage caused by implantation, and smoothing surfaces for the deposition of the second dielectric. An increase in fixed charge, $Q_{fb}$, and interface trap densities, $D_{it}$, of $10^{11}$ cm$^{-2}$ and $10^{10}$ cm$^{-2}$ eV$^{-1}$, respectively, were found in samples that witnessed an equivalent Si implantation dose, energy and a 24 minute 900°C reoxidation [10]. Under this thermal budget, we estimate a 5 nm diffusion length for Sb, which is small relative to both the implant depth and lateral implant window size. Finally, 60 nm of aluminum oxide (Al$_2$O$_3$) is deposited using atomic layer deposition (ALD) that isolates the poly-Si gates from the Al top gate. The Al top gate covers the entire sample and is biased positively to enhance an inversion layer of electrons at the Si/SiO$_2$ interface while the poly-Si gates are biased negatively to locally deplete electrons thus creating a single tunnel barrier. A cross-section schematic of a completed Sb implanted device is shown in figure 1(b). A thorough description and study of the fabrication process flow of similar devices is presented in reference [5] although the use of implant and oxidation for annealing of the implant are specific to this work.

After fabrication, we measure the differential conductance's dependence on gate and source-drain bias when immersed in liquid He. We measure the differential conductance $G = dI/dV_{SD}$ through the split gate structure using standard lock-in techniques with an rms ac signal of 100 μV added to a dc source-drain bias $V_{SD}$. Figure 1(c) and (d) show the differential conductance as a function of the Al top gate voltage $V_{AG}$ and $V_{SD}$ at a constant TP and CP voltage for samples A and B, respectively. The non-implanted PC shows no evidence of resonances indicating that the disorder potential is small relative to the thermal energy $k_B T$ in this particular bias range. The conductance dependence on voltage qualitatively mimics clean behavior in the more ideal GaAs split gate configuration [1,2] showing regions of strong dependence on either $V_{SD}$ (positive $V_{AG}$) or $V_{AG}$ (large magnitude $V_{SD}$) but without evidence of ballistic transport at this temperature. Transport through the Sb implanted PC, figure 1 (d), shows additional resonances and a significant positive offset in threshold voltage. This case illustrates the effect on transport when local potentials are introduced in a tunnel barrier. The resonant transport is qualitatively consistent with several other reports of shallow donor levels that are created in tunnel barriers after donors are introduced into the tunnel barrier region, for example As [3] and P [4]. To our knowledge this is the first report of clean transport spectroscopy of an Sb implanted tunnel barrier.

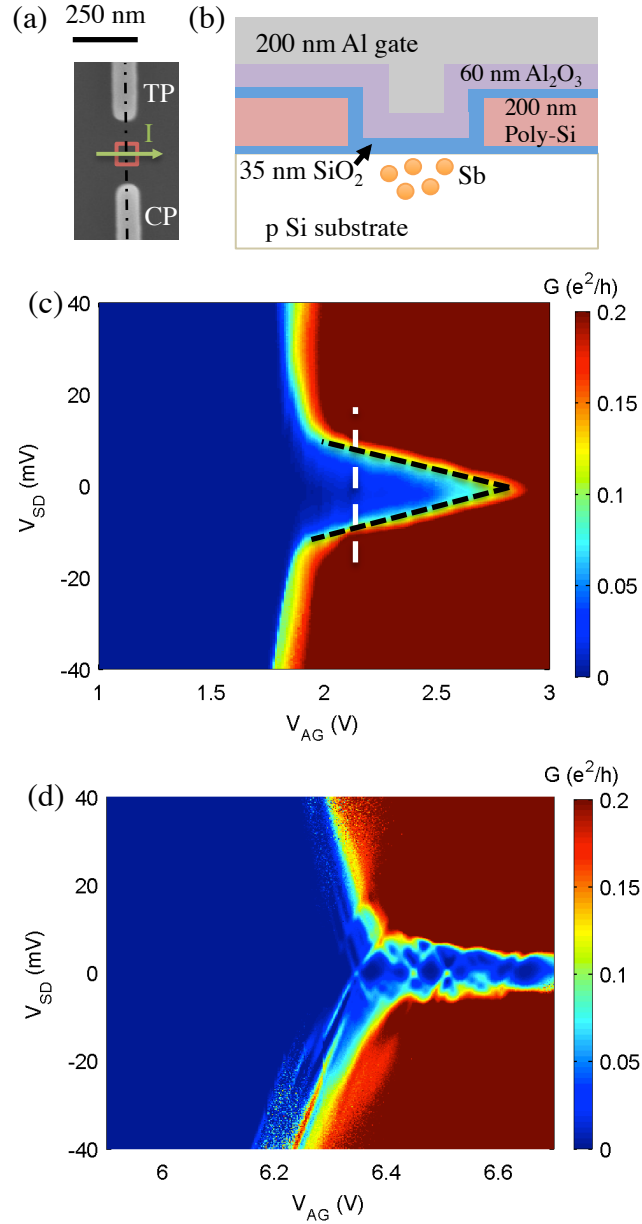

**Figure 1.** (a) SEM image of a Si split gate structure before deposition of the $Al_2O_3$ layer and Al top gate. The square represents the 80 nm × 80 nm implantation window for the Sb donors and the arrow represents the direction of current. (b) (Not to scale) Cross-sectional schematic diagram of a completed device along the dash-dotted line in panel a. (c) $dI/dV_{SD}$ as a function of $V_{AG}$ and $V_{SD}$ for the non-implanted PC (Sample A). TP and CP are held constant at $V_{TP} = V_{CP} = -2.5$ V. The white dashed line indicates the direction along which line fits were made to extract barrier height and width. The black dashed line is a guide to the eye to indicate a contour of constant conductance discussed in the text. (d) $dI/dV_{SD}$ as a function $V_{AG}$ and $V_{SD}$ for an Sb-implanted PC (Sample B) with $V_{TP} = V_{CP} = -4.85$ V.

We model the tunnel barrier using a rectangular barrier model [11] that depends on the 1D parameters width and barrier height:

$$I \approx \left(3\sqrt{2mU}/2w\right)(q/h)^2 A V_{SD} \exp\left(-\frac{2w}{\hbar}\sqrt{2mU}\right) \quad (1)$$

Here $U$ is the barrier height, $w$ is the width, $m$ is the effective mass of an electron, $q$ is the electron charge, $h$ is Planck's constant, and $A$ is the cross-sectional area perpendicular to the transport direction. The voltage dependence of the current is modeled by starting with MacLean et al.'s model that the tunnel barrier height depends linearly on $V_{SD}$ and $V_{AG}$ [7],

$$U = U_0 - q\alpha_{SD} V_{SD} - q\alpha_{GATE} V_{GATE} \quad (2)$$

Where $U_0$ is the barrier height at zero applied bias and $\alpha_{SD}$ and $\alpha_{GATE}$ are the capacitive couplings to the source/drain leads and the gate, respectively. In our case, $V_{GATE} = V_{AG}$. To estimate the values of $\alpha_{SD}$ and $\alpha_{GATE}$, we assume that the energy difference between the top of the barrier and the Fermi level are constant for the linear contours of constant conductance in the un-implanted barrier (i.e, $V_{AG} > 2V$) and implanted barrier (i.e., $V_{AG} > 6.4V$), figure 1(c) black dashed lines. That is, a small change in the tunnel barrier height, $dU$, is linearly dependent on small changes in the dc source-drain bias $dV_{SD}$ and the gate voltage $dV_{GATE}$. These are balanced for any contour of constant conductance in the linear region indicated by the black dashed lines. The positive and negative slopes can then be associated with $m_p = C_{GATE}/(C - C_S)$ and $m_n = C_{GATE}/C_S$, respectively. Here $C$ is the total capacitance between all conductors and the barrier, $C_S$ is the capacitance between the source and the barrier, and $C_{GATE}$ is the capacitance between the gate and the barrier. This method of extracting the capacitance ratios is similar for quantum dots [12]. Then $\alpha_{SD}$ and $\alpha_{GATE}$ are:

$$\alpha_{SD} = \frac{C_S}{C} = \frac{1}{1 + m_n/m_p} \quad (3)$$

$$\alpha_{GATE} = \frac{C_{GATE}}{C} = \frac{1}{\frac{1}{m_n} + \frac{1}{m_p}} \quad (4)$$

We substitute (2) into (1) and use a Taylor series expansion for the square root term similar to reference [7] to approximate the current as $I \approx I_0 \exp(\beta V_{SD})$ for constant values of $V_{AG}$. Here $I_0$ is the intercept and $\beta$ is the slope of the region where the current turns on exponentially (see figure 2(a)). Both $I_0$ and $\beta$ depend on the $U$ and $w$.

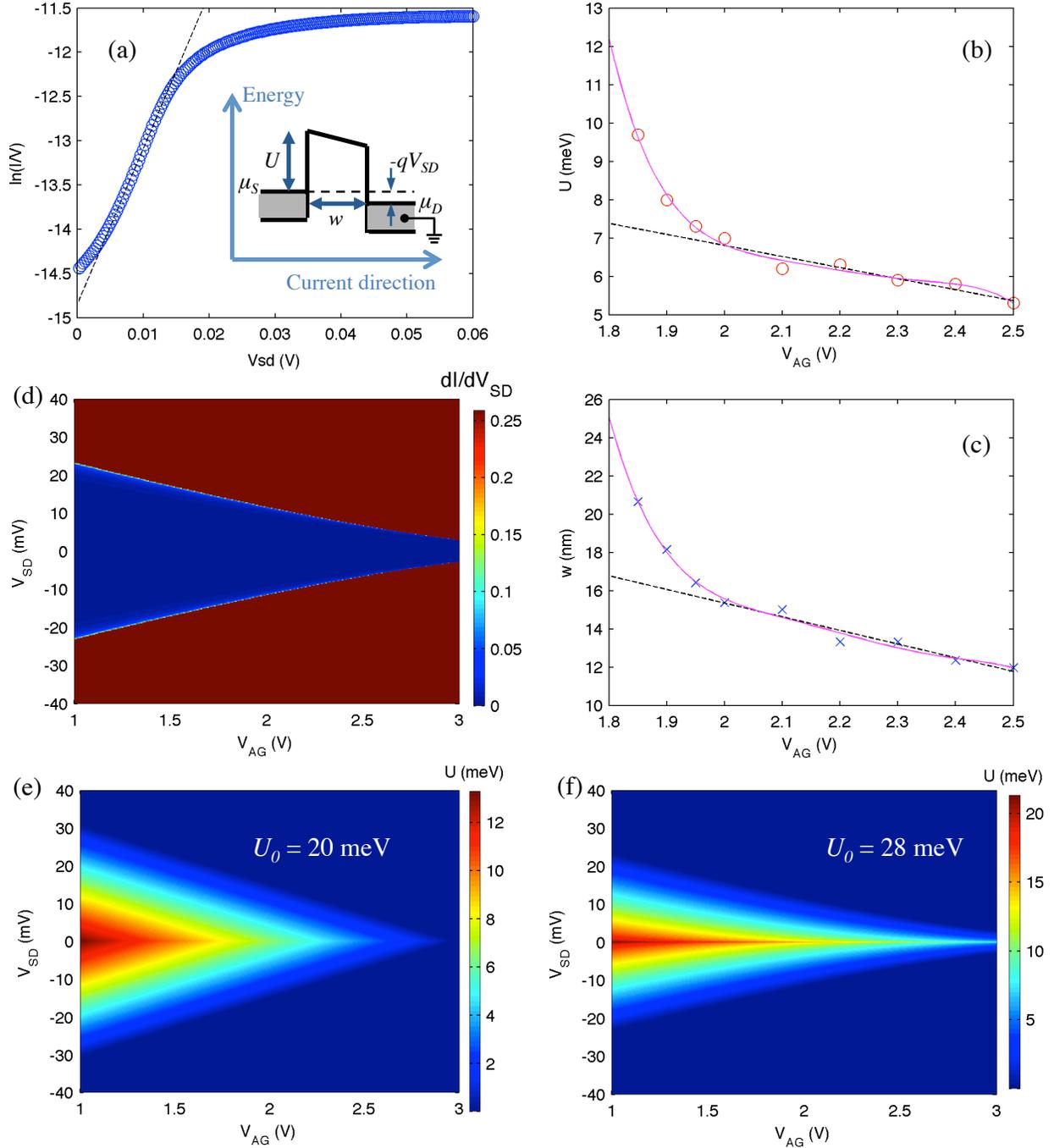

**Figure 2.** (a) Line cut along $V_{AG} = 2.1$ V from figure 1(c), showing an exponential increase in the current through the point contact as $V_{SD}$ increases. The parameters $I_0$ and $\beta$ are found from the y-intercept and slope of the line fit (black dashed line) to the exponential region, respectively. Inset shows a schematic energy diagram of the tunnel barrier and definitions of $U$ and $w$. The chemical potentials of the source and drain reservoirs are given by $\mu_S$ and $\mu_D$, respectively. (b) $U$ dependence on $V_{AG}$ extracted for $V_{SD} = 0$. A linear fit can be used to approximate $U$ at high $V_{AG}$ while the curvature at low $V_{AG}$ can be fit by a higher order polynomial. This is similar for $w$ shown in panel (c). (d) d$I$/d$V_{SD}$ simulation as a function of $V_{SD}$ and $V_{AG}$. The simulation uses a linear dependence of $U$ and $w$ on $V_{SD}$ and $V_{AG}$ and also accounts for width

narrowing and image barrier lowering at high $V_{SD}$. (e) Simulated $U$ without image barrier lowering. (f) Simulated $U$ with image barrier lowering.

## 3. Results

We extract $I_0$ and $\beta$ from $V_{SD}$ line cuts at different fixed $V_{AG}$, an example at $V_{AG}$ = 2.1V for the non-implanted barrier is shown in figure 2(a). The line cuts are used to establish the barrier heights and widths at each voltage $V_{AG}$ (figures 2 (b) and (c)). We estimate the cross-sectional area $A$ as the geometrical length for the barrier multiplied by the depth of the channel. From the SEM image in figure 1(a) the distance between the tips of the plunger gates is 240 nm but the ALD $Al_2O_3$ fills 60 nm on other side of the constriction. Thus we estimate the length of the barrier to be 120 nm and the channel to be 5 nm deep [13,14]. The area of the point contact is a pre-factor in the current so inaccuracy in the choice of area will have a relatively small effect on the extracted barrier heights and widths shown in figure 2. We observe a linear dependence of $w$ and $U$ on $V_{AG}$ for a wide range of gate bias. A similar linear dependence as well as similar magnitudes of $U$ and $w$ were predicted from a more complete semi-classical numeric simulation of the point contact geometry by Gao et al., although the dependence was calculated for a plunger gate not the top Al gate [15]. The linear dependence of the barrier height on $V_{AG}$ is also qualitatively self-consistent with the MacLean model (i.e., the starting assumption for the model). The quantitative agreement can be examined by comparing the extracted dependence of the barrier height on voltage from figures 2(b) and (c) to what is predicted by $\alpha_{SD}$ and $\alpha_{GATE}$ obtained from our capacitance model. We find a discrepancy of 2 meV or less in barrier height. This is the dominant source of numeric uncertainty in the rectangular barrier height. We expect disagreement because the initial estimate from the capacitance model uses a fixed width assumption.

The linear model fails to predict the tunnel barrier behavior at less positive enhancement gate bias. Fowler-Nordheim (FN) width narrowing [11] and image charge barrier lowering [16] (ICBL) are both mechanisms that affect current at high lateral fields. Direct fitting of the functional form of the constant conductance contours to the standard forms of these models, however, does not produce the observed current dependence on gate bias in the more negative bias regime using a standard FN form [11] or the following common approximation for ICBL, $\Delta U = \sqrt{q^3 F / 4\pi\varepsilon_S}$ where $F$ is the electric field due to the source-drain bias and $\varepsilon_S$ is the electric permittivity of the semiconductor [16]. An example simulation using linear dependence of the barrier height and width on voltage, combined with high field width narrowing and ICBL effects is shown in figure 2(d). The model clearly fails to predict the more complex contours of constant conductance seen in the experiment (figure 1). The simulation with ICBL does highlight that to produce similar conductance values requires, consequently, larger barrier heights in the low-field regime for the same gate biases. That is, the rectangular barrier height fits are a lower estimate and inclusion of ICBL leads to roughly an 8 meV increase in barrier height for this geometry and bias range as shown in figures 2(e) and (f).

To describe the non-linear part of the contours of conductance or funnel-like behavior in the simulation, we extend the rectangular barrier height and width fits to these regions (i.e., $V_{AG}$ < 2V for the non-implanted case and $V_{AG}$ < 6.4V for the implanted case). Barrier heights and widths are shown in figures 2(b) and (c). Empirically, we find that the barrier height and width dependence on gate voltage is non-linear. The dependence can be fit to a polynomial dependence on voltage. From this empirical fit, this region can be simulated including the high field effects and we find that the simulation qualitatively captures the funnel-like behavior, as shown in figure 3(a). The conductance is also quantitatively within an order of magnitude of the experiment. This is done by using the polynomial fit and with no modification of further fitting parameters. The improved estimates of barrier heights and widths in this bias region are shown in figures 3(b) and (c). We use a similar analysis to simulate the differential conductance, barrier height, and width for the implanted tunnel barrier, figures 3(d-f).

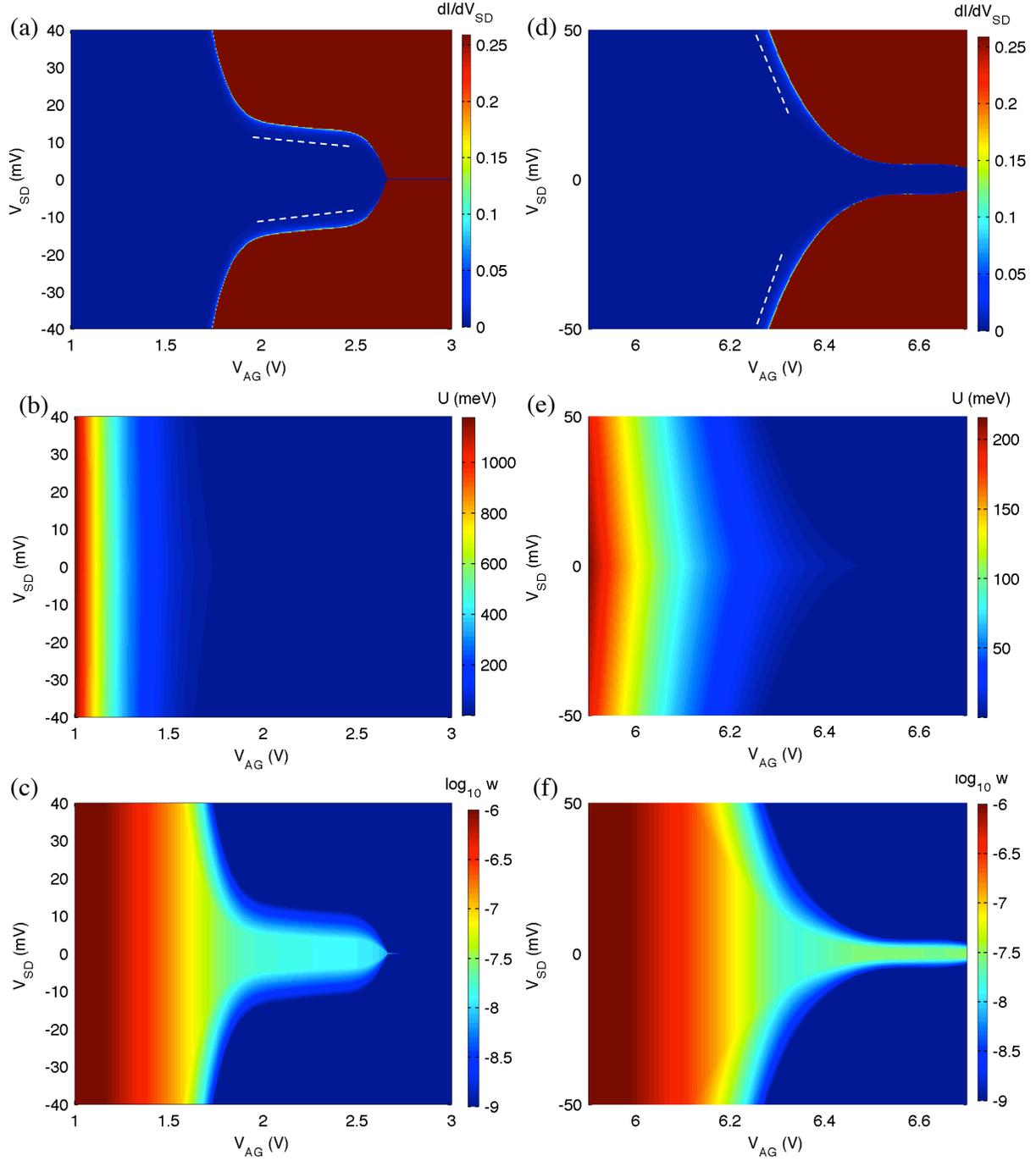

**Figure 3.** (a) d$I$/d$V_{SD}$ simulation of the non-implanted barrier using a non-linear model for $U$ and $w$. White dashed line is a guide for the eye to show where the conductance first turns on. (b) Simulated $U$ for the non-implanted case. (c) Simulated $w$ for the non-implanted case. (d) d$I$/d$V_{SD}$ simulation of the Sb-implanted barrier using a non-linear model for $U$ and $w$. White dashed line is a guide for the eye to show where the conductance first turns on. (e) Simulated $U$ for the Sb-implanted case. (f) Simulated $w$ for the Sb-implanted case.

## 4. Discussion

Using this model, we can estimate the barrier heights and widths at each location for which a resonance is observed. We find that the last resonance seen at low gate voltage ($V_{AG}$ ~ 6.35V) has a barrier height of approximately 12 meV, while the barrier heights of the subsequent resonances decrease as the gate voltage increases with values ranging from 6 to 2 meV. These values increase when we include image-charge barrier lowering to 20 meV for the last resonance and 13 to 10 meV for the resonances at higher gate voltage. This suggests that the last resonance at low $V_{AG}$ regardless of model choice has a relatively large binding energy and thus is more consistent with a $D^0$ transition, implanted donor. The additional resonances observed in the implanted barrier case, figure 1(d), are consistent with previous reports of transport through one or several donors in the tunnel barrier [4]. Binding energies below the bulk value are expected and have been reported for other impurities in Si [3]. The resonances at higher $V_{AG}$ are due to unidentified states with shallower binding energies. The identification is more ambiguous since the $D^-$ transition, shallow traps, or defect states from process induced damage at the MOS interface are all possible candidates that have binding energies in the 1-5 meV range [17,18] We note that binding energy extraction in many other donor transport spectroscopy studies has relied on more indirect approaches, for example assignment of $D^0$ and $D^-$ transitions from similar slopes and B-field spin filling dependences, which can be used to extract a charging energy and utilized as a measure of the binding energy. This approach can leave doubt, particularly in tunnel junctions with many donors that have many resonant lines many of which can also have similar capacitances to the gates making it sometimes very difficult to make assignments necessary to extract a charging energy.

## 5. Conclusion

In summary, we fabricated MOS split-gate tunnel barriers with and without Sb implants and performed transport spectroscopy at ~4K. The un-implanted Si MOS split gate conductance is free of resonant behavior at this temperature and shows contours of constant conductance that are qualitatively similar to those observed in split gate point contacts from other more model material systems (e.g., GaAs). The implanted point contact in contrast has resonances in the transport spectroscopy. The resonant behavior is consistent with previous reports of transport spectroscopy of P or As in Si tunnel barriers for which the resonances are associated with transport through single donors. These tunnel barriers are distinguished from previous experimental reports because of (a) the specific fabrication process that uses a split-gate geometry, poly-silicon conductors and oxidation to anneal damage combined with (b) the use of Sb as the implanted donor for which, to our knowledge, there are no other reports of clear transport spectroscopy in the literature.

The barrier height and width dependences on voltages are subsequently modeled with a modified 1D rectangular barrier model. The model considers both shifting of the barrier height relative to the Fermi level due to the applied gate voltages and high field effects due to lateral fields applied across the source-drain contacts, such as Fowler-Nordheim like barrier narrowing. Barrier heights and widths can be estimated from this simulation and are quantitatively similar to semi-classical numeric simulations of related point contact geometries. Reasonable agreement with the measured conductance is also observed. Using this method, we estimate the barrier heights in the location of the resonances in the implanted tunnel barrier, which provides estimates of the depth of the resonant levels below the top of the barrier. A few of the resonances are relatively deep (~20 meV) and are consistent with transport through Sb atoms near the surface.

**Acknowledgements**

We gratefully recognize conversations with Dr. J. Gamble, Dr. R. Lewis and Dr. T-M. Lu about this work. This work was performed, in part, at the Center for Integrated Nanotechnologies, a U.S. DOE, Office of Basic Energy Sciences user facility. The work was supported by the Sandia National